\documentclass[aps,prd,preprint,nofootinbib,superscriptaddress]{revtex4-1}

\usepackage{amsmath,amssymb,graphicx}
\usepackage{tabularx}
\newcolumntype{C}[1]{>{\centering\arraybackslash}p{#1}}
\usepackage{hyperref}

\usepackage{xcolor}

\begin{document}
\author{Alessandro Casalino}
\email{alessandro.casalino@unitn.it}
\affiliation{Dipartimento di Fisica, Universit\`{a} di Trento,\\Via Sommarive 14, I-38123 Povo (TN), Italy}
\affiliation{Trento Institute for Fundamental Physics and Applications (TIFPA)-INFN,\\Via Sommarive 14, I-38123 Povo (TN), Italy}
\author{Lorenzo Sebastiani}
\email{lorenzo.sebastiani@pi.infn.it}
\affiliation{Istituto Nazionale di Fisica Nucleare, Sezione di Pisa, Italy}
\affiliation{Dipartimento di Fisica, Universit\'a di Pisa, Largo B. Pontecorvo 3, 56127 Pisa, Italy}
\title{Perturbations in Regularized Lovelock Gravity}

\begin{abstract}
In this paper we study the perturbation theory of the recently proposed Regularized Lovelock Gravity \cite{Glavan:2019inb}, on the curved Friedmann-–Lema\^{i}tre-–Robertson-–Walker (FLRW) space-time. We provide the first order perturbation equations both in the scalar and tensor sector in the presence of an additional minimally coupled scalar field. A general expression for the velocity of gravitational waves at the generic order is inferred. Moreover, we apply the results on the study of slow-roll inflation on flat FLRW background, at second order in the Regularized Lovelock Gravity, i.e. in the presence of the Gauss-Bonnet correction only. We derive the power spectra of scalar and tensor perturbations, and provide the equations for the inflation observable quantities: the spectral indices and the tensor-to-scalar spectra ratio.
\end{abstract}

\allowdisplaybreaks[1]

\maketitle

\section{Introduction}

The Lovelock theorem \cite{Lovelock:1971vz} provides the most general gravitational theory, described by the Lovelock-Lanczos action, leading to second order field equations in $d$ dimensions. In four dimension, since higher orders Lovelock terms contribute to the field equations with identically vanishing contributions, the theory reduces to General Relativity (GR) with a cosmological constant \cite{Lovelock:1972vz}.

There are different ways to circumvent the Lovelock theorem in the metric formalism, without adding degrees of freedom to the action. For instance, one method considers theories leading to second order field equations only for specific classes of metric fields. This is usually exploited in Quasi-Topological gravity theories \cite{Quasi-Top1, Quasi-Top2} and Non-polynomial gravity theories \cite{Deser, LorenzoSergio, aim}. 

In this paper we will consider a different approach, firstly proposed in the four dimensional case by Tomozawa \cite{Tomozawa:2011gp}. Consider the following regularized Einstein-Gauss-Bonnet theory
\begin{equation}
S = \int d^d x \sqrt{-g} \left( \frac{M_P^2}{2}R + \frac{\alpha}{d-4} \mathcal{G} \right)\,,
\end{equation}
where $\mathcal{G} \equiv R^2 - 4 R^{\mu \nu} R_{\mu \nu} + R^{\mu \nu \rho \sigma} R_{\mu \nu \rho \sigma}$ and $M_p^2 \equiv 1/8 \pi G$. Tomozawa proved that, if the associated equations of motion are evaluated considering a static spherically symmetric ansatz, we can set the dimension to $d=4$, preserving the contribution coming from the Gauss-Bonnet term. The resulting theory admits a black hole solution with a repulsive gravitational force near the time-like singularity $r\to 0$. Note that the same solution has been found in different contexts motivated by quantum corrections to gravity. See for instance the Ho\v{r}ava-Lifshitz gravity case \cite{Horava}, and the semi-classical Einstein equations with conformal anomaly \cite{conformal}. The geodesic structure, as well as its quasi-normal modes and stability, have been investigated recently in, respectively, \cite{geodesics} and \cite{QNM}. The charged case was investigated in \cite{conformal} and more recently in \cite{charged}. The Anti de Sitter (AdS) black holes and their thermodynamics were found and discussed in \cite{ConfThermo, thermod}. Finally, a rotating generalization has been studied in \cite{Rotating,Rotating2}. Other works on black holes in the context of this theory can be found in \cite{bh1,bh2,bh3,bh4,bh5,bh6,bh7,bh8,bh9,bh10,bh11,bh12,bh13,bh14,bh15,bh16, bh17, bh18}.
In \cite{LiLi} the authors study the emergent universe scenario in Regularized Lovelock Gravity, and obtain the stability of Einstein static universe against both homogeneous and inhomogeneous scalar perturbations simultaneously. In \cite{Shu} the authors study the instabilities of the theory. Moreover, the role of the Gauss-Bonnet term in four dimensional gravity and in particular holographic
implications to the addition of it to the gravity action was studied in \cite{rodrigo}.

The same theory has been also studied by Cognola et al. \cite{Cognola:2013fva}. They showed that the entropy of the black hole is logarithmic in its area, which provides a motivation for interpreting these terms as quantum corrections. In fact, such a behaviour of the entropy typically arises once the quantum effects are taken into account \cite{Carlip,PerezNoui}. Finally they proved that the flat FLRW sector of this theory is well-defined in the limit $d\to 4$. An analogous investigation in the case of spatially curved FLRW sector has also been recently studied \cite{Casalino:2020kbt}.

Glavan and Lin further extended the applicability of this theory to first order in perturbation theory around (A)dS$_4$ vacua. In fact, they showed that this theory only contains the degrees of freedom of a massless graviton, as in GR \cite{Glavan:2019inb}, which is a very crucial result to assess the viability of this theory, as it indicates that it is not plagued by ghost-like instabilities. Although this is not a sufficient proof, it provides a strong indication that the regularization of Lovelock-Lanczos gravity is a well-defined procedure.

Recently, a discussion arose about the well-posedness of this regularization. See for instance \cite{Gurses:2020ofy,Ai:2020peo}. In fact, at least for FLRW or SSS metric fields, is always possible to extract the dimensional factor from the equations of motion. In other words, for FLRW or SSS metric fields $g_{\rho \sigma}$, is always possible to write the equations of motion in the form
\begin{equation}
    G_{\mu \nu} \left[g_{\rho \sigma}\right] = (d-D) H_{\mu \nu} \left[g_{\rho \sigma}\right]\,,
\end{equation}
where $H_{\mu \nu}$ is a well-defined covariant tensor, and the limit $d\rightarrow D$ is possible for any $D$. This, for a generic metric, is not possible, creating questions on the well-posedness of the whole regularization procedure. However, in this paper we limit our discussion to the FLRW metric, which is not affected by this problem, and use the theory with its regularization as an effective theory with a prescription to obtain the equations of motion with non trivial corrections with respect to General Relativity. This has also been noted by \cite{Malafarina:2020pvl}. On the subject, see also \cite{Mahapatra:2020rds}.

Finally, a reformulation of Regularized Lovelock theory with a sub-class of Horndeski gravity has been recently proposed \cite{horndeski1,horndeski2}.

In this paper we will extend the results obtained by Glavan and Lin \cite{Glavan:2019inb}, and by Casalino et al \cite{Casalino:2020kbt}. In particular we will study the first order perturbation theory in the case of curved FLRW with an additional minimally coupled scalar field, providing the perturbations equation both in the scalar and tensor sector. In particular, the equations are derived at third order in Regularized Lovelock Gravity corrections, and then we infer a generalization for any Lovelock order. We will then derive the formalism of first order perturbation theory for inflation in the reduced case of regularized Gauss-Bonnet theory (second order of Regularized Lovelock Gravity), which is the second order in Regularized Lovelock Gravity, in a flat FLRW.

The paper is structured as follows. In Sec. \ref{sec:regul_love_theory} we provide a brief introduction to the Regularized Lovelock Gravity theory, showing the associated regularization and providing the general equations of motion. In Sec. \ref{sec:back} we show the background results on a spatially curved FLRW background. In the following sections Sec. \ref{sec:tens_pert} and \ref{sec:sc_pert}, we derive respectively the tensor and scalar perturbation equations. We also provide the equations for the tensor and scalar waves velocity. In Sec.\ref{sec:inflation} we study the corrections to slow-roll scalar field inflation given by Lovelock theory. Finally, in Sec. \ref{sec:conclusions} we draw the conclusions.

In this paper we follow the convention $c=1$. Moreover, we define the Planck mass as $M_p^2 \equiv 1/8 \pi G_N$.

\section{Regularized Lovelock Gravity}\label{sec:regul_love_theory}

In this section we review the Regularized Lovelock Gravity theory. We consider the following $d-$dimensional action
\begin{equation}
    S = \int d^dx \,\sqrt{-g} \, \left[\sum_{p=0}^t \alpha_p \, \mathcal{L}_p\left(R,R_{\mu \nu},R_{\mu \nu \rho \sigma}\right)\right]\, + S_\varphi\,,\label{eq:action_g}
\end{equation}
where the first part is the Lovelock--Lanczos action, while the second $S_\varphi$ part is the contribution from a scalar field which is minimally coupled to the metric. The parameter $t$ is the order of the Lovelock gravity, $\alpha_p$ the coupling constants and $\mathcal{L}_p$ are curvature tensor functions, defined as
\begin{equation}
\mathcal{L}_p = \frac{1}{2^p} \delta^{\mu_1 \nu_1 \dots \mu_p \nu_p}_{\sigma_1 \rho_1 \dots \sigma_p \rho_p} \prod  _{r=1}^p R^{\mu_r \nu_r}_{\phantom{\mu_r \nu_r} \sigma_r \rho_r}\,,
\end{equation}
where $\delta^{\mu_1 \nu_1 \dots \mu_p \nu_p}_{\sigma_1 \rho_1 \dots \sigma_p \rho_p}$ is the generalized Kronecker delta. Explicitly, the definition of these functions up to the third order in Lovelock gravity are
\begin{align}
\mathcal{L}_0 &= 1 \,,\label{eq:L_0}\\
\mathcal{L}_1 &= R \,,\label{eq:L_1}\\
\mathcal{L}_2 &= R^2 - 4 R^{\mu \nu} R_{\mu \nu} + R^{\mu \nu \rho \sigma} R_{\mu \nu \rho \sigma} \equiv \mathcal{G}\,, \label{eq:L_2}\\
\mathcal{L}_3 &= R^3   -12RR_{\mu \nu } R^{\mu \nu } + 16R_{\mu \nu }R^{\mu }_{\phantom{\mu } \rho }R^{\nu \rho }+ 24 R_{\mu \nu }R_{\rho \sigma }R^{\mu \rho \nu \sigma }+ 3RR_{\mu \nu \rho \sigma } R^{\mu \nu \rho \sigma } \nonumber \\
&-24R_{\mu \nu }R^\mu _{\phantom{\mu } \rho \sigma \kappa } R^{\nu \rho \sigma \kappa  }+ 4 R_{\mu \nu \rho \sigma }R^{\mu \nu \eta \zeta } R^{\rho \sigma }_{\phantom{\rho \sigma } \eta \zeta }-8R_{\mu \rho \nu \sigma } R^{\mu  \phantom{\eta } \nu }_{\phantom{\mu } \eta 
\phantom{\nu } \zeta } R^{\rho  \eta  \sigma  \zeta }\,. \label{eq:L_3}
\end{align}

Moreover, we define the scalar field part of the action as
\begin{equation}
    S_\varphi = \int d^d x \, \sqrt{-g} \left[ - \frac{1}{2}\nabla_\mu \varphi \nabla^\mu \varphi - V(\varphi)\right]\,,\label{eq:action_scalarfield}
\end{equation}
where the potential $V(\varphi)$ is a scalar function of the field $\varphi$. Note that, since the scalar field is minimally coupled to the metric, the contribution from $S_\varphi$ to the equations of motion will be a simple addition.

The equations of motion can be computed varying the action in Eq. \eqref{eq:action_g} respectively with respect to the metric $g_{\mu \nu}$ and $\varphi$. We obtain
\begin{align}
\mathcal{G}_{\mu \nu} = \sum_{p=0}^t \alpha_p \, \mathcal{G}^{(p)}_{\mu\nu} &= \frac{1}{2}\, \left[\left(\nabla_\mu \varphi \right) \left(\nabla_\nu \varphi\right) - \frac{1}{2}g_{\mu\nu} \left(\nabla_\rho \varphi \right) \left(\nabla^\rho \varphi\right) - g_{\mu\nu} V(\varphi)\right],
\label{eq:eom_t}\\
\nabla_\mu \nabla^\mu \varphi - V(\varphi) &= 0\,, \label{eq:eom_scalarfield}
\end{align}
where 
\begin{equation}
\label{eq:Lovelocktensor}
\mathcal{G}_\beta^{(p) \alpha}\big[g_{\mu\nu} \big]  = \frac{g^{\alpha\gamma}}{\sqrt{-g}}\frac{ \delta\left[\sqrt{-g} \mathcal{L}_p \right]}{\delta g^{\gamma\beta}} =  -\frac{1}{2^{p+1}} \delta^{\alpha \mu_1 \nu_1 \dots \mu_p \nu_p}_{\beta \sigma_1 \rho_1 \dots \sigma_p \rho_p}\prod  _{r=1}^p R^{\mu_r \nu_r}_{\phantom{\mu_r \nu_r} \sigma_r \rho_r} \, .
\end{equation} 
Since in the equations of motion \eqref{eq:Lovelocktensor} we have the totally anti-symmetric Kronecker delta, the $d \leq 2p$ contributions identically vanish. To circumvent this problem, we can implement the aforementioned regularization by re-scaling the coupling constants $\alpha_p$, and then by fixing the dimension of the manifold. With this procedure we obtain non-vanishing contributions to the equations of motion coming from generic Lovelock gravity orders. Therefore, we define the following coupling constants
\begin{equation}
    \alpha_1 = \frac{M_P^{d-2}}{2} \qquad \alpha_2 = \frac{\alpha}{d-4} \qquad \alpha_3 = \frac{\beta}{d-4}\,,
\end{equation}
where the $d-4$ in the denominator has been added to implement the aforementioned regularization. In our discussion, we will consider a vanishing cosmological constant contribution, i.e. $\alpha_0 \propto \Lambda = 0$.

In the following we will consider these equations of motion on a Friedmann-–Lema\^{i}tre-–Robertson-–Walker (FLRW) manifold.

\section{Background on FLRW}\label{sec:back}

In this section we review some results of Lovelock gravity on a FLRW manifold. These results were derived in \cite{Glavan:2019inb}, and generalized in \cite{Casalino:2020kbt}. Here we present a rewrite of the equations in a more compact form. We consider the metric
\begin{equation}
    ds^2 = - dt^2 + a(t)^2 g_{ij}^{(d-1)}(x^k)\, dx^i dx^j\,,
\end{equation}
where $g_{ij}^{(d-1)}$ is the spatially curved $d-1$ dimensional metric, with curvature constant $\kappa$, written in some spatial coordinates $x^i$. Using the equations of motion \eqref{eq:eom_t} and \eqref{eq:eom_scalarfield}, we obtain the equations of motion
\begin{align}
    F(J^2)&=\left[\frac{\dot \varphi^2}{2} + V(\varphi)\right]\,,\label{eq:motion0}\\
    (Q^2-J^2)F'(J^2)&=-\dot \varphi^2\,,\label{eq:motion}\\
    \ddot \varphi + (d-1) H \dot \varphi + V(\varphi) &= 0\,,\label{eq:KG0}
\end{align}
where we define the functions $J^2$ and $Q^2$ as
\begin{equation}
    J^2 \equiv H^2 + \frac{\kappa}{a^2} \qquad \text{and} \qquad Q^2 \equiv H^2 + \dot{H}\,,
\end{equation}
The function $F$ is defined as
\begin{equation}
    F(J^2)\equiv \sum_{p=1}^{\infty} \left[\prod_{s=1}^{2p}(d-s)\right]\alpha_p \,J^{2p}\,,
\end{equation} 
and its derivative reads
\begin{equation}
    F'(J^2)\equiv \frac{d F(J^2)}{d J^2}\,.
\end{equation}

In particular, at cubic order in Lovelock gravity ($t=3$), in the limit $d \rightarrow 4$, we obtain
\begin{align}
    J^2 \left(1 + \frac{2 \alpha}{M_P^2}J^2 + \frac{4 \beta}{M_P^2}J^4 \right) &= \frac{1}{3 M_P^2} \left[\frac{\dot \varphi^2}{2} + V(\varphi)\right]\,,\label{eq:friedm1}\\
    \left(Q^2-J^2\right) \left(1+\frac{4 \alpha}{M_p^2} J^2 + \frac{12 \beta}{M_p^2} J^4 \right)&= -\frac{\dot \varphi^2}{2 M_p^2}\,,\label{eq:friedm2}\\
    \ddot \varphi +3 H \dot \varphi + V(\varphi) &= 0\label{eq:KG}\,.
\end{align}

\section{Tensor perturbations}\label{sec:tens_pert}

In this section we present the first order tensor perturbation equations in a curved FLRW space. We consider the following line element
\begin{equation}
    ds^2 = - dt^2 + a(t)^2 \left[ g_{ij}^{(d-1)}(x^k) + {h_{ij}}^{(d-1)}(t,x^k)\right] dx^i dx^j\,,
\end{equation}
where $h$ is a $d-1$ dimensional spatial tensor whose value is always less than unity. The following results are shown in the limit $d \rightarrow 4$.

In order to obtain the tensor waves perturbation equation, we perform the computation on the $i-j$ Einstein equation with $i \neq j$. We obtain
\begin{align}
    \ddot{h}_{ij} &+ 3 H \left[1 + \frac{8}{3 \Gamma M_{P}^2} \left(\alpha + 6 \beta J^2\right) (Q^2-J^2) \right] \dot{h}_{ij} \nonumber\\
    &+ \left[1+\frac{8}{ \Gamma M_{P}^2} \left(\alpha + 6 \beta J^2\right) (Q^2-J^2) \right]\left(\frac{2 \kappa}{a^2} - \frac{\partial_{\, l} \partial^{\,l}}{a^2}\right)h_{ij}=0\,,
\label{tp}
\end{align}
where $Q^2-J^2=\dot H - \kappa/a^2$ by definition, and $\Gamma$ is a dimensionless factor defined as
\begin{equation}
    \Gamma \equiv 1 + \frac{2}{M_{P}^2} \left(2\alpha J^2 + 6\beta J^4 \right)\,.
\end{equation}
In our next analysis of the perturbation theory this factor will play an important role.
In the limit of General Relativity $\Gamma=1$, and the deviations from one depend on the magnitude of Lovelock corrections (here written up to the third order) with respect to the Planck Mass squared. 

Therefore we can define the velocity of gravity waves as
\begin{equation}
    c_T^2 = 1+\frac{8}{\Gamma M_{P}^2} \left(\alpha + 6 \beta J^2\right) (Q^2-J^2)\,,\label{eq:ct2}
\end{equation}
from which we can see that the addition of the regularized Gauss-Bonnet adds a correction to the General Relativity result $c_T^2=1$. In an expanding universe dominated by matter $\Gamma\simeq 1$, and the second term in Eq. \eqref{eq:ct2} is strongly suppressed by the Planck Mass squared. On the other side, when the curvature of the universe is non-negligible with respect to the Planck Mass squared value, the $1/\Gamma$ factor is larger than one and the correction to the General Relativity result $c_T^2=1$ can still be taken arbitrarily small.

We can also infer a generalization of the tensor perturbation equations \eqref{tp} to a generic order $t$ in the Lovelock series as
\begin{align}
   \ddot{h}_{ij} + 3 H \left(1 + \frac{\Theta_t}{3} \right) \dot{h}_{ij} +\left(1 + \Theta_t \right)\left(\frac{2 \kappa}{a^2} - \frac{\partial_{\, l} \partial^{\,l}}{a^2}\right)h_{ij}=0 \,,
\end{align}
where we define the adimensional coefficients
\begin{equation}
    \Theta_t \equiv \frac{Q^2-J^2}{\Gamma_t M_{P}^2} \sum_{p=2}^{p=t} 2^p p! \, \alpha_p J^{2(p-2)}\,\label{eq:Theta}
\end{equation}
and
\begin{equation}
    \Gamma_t \equiv \frac{2}{M_{P}^2} \sum_{p=1}^{p=t} p! \, \alpha_p J^{2(p-1)} = 1 + \frac{2}{M_{P}^2} \sum_{p=2}^{p=t} p! \,\alpha_p J^{2(p-1)}\,.
\end{equation}
Note that when $t=1$, which is the General Relativity case, $\Gamma_t = 1$ and $\Theta_t = 0$.

Finally, we can write the velocity of gravity waves up to order $t$ in the Lovelock series
\begin{equation}
    c_T^2 = 1 + \Theta_t\,.
\end{equation}

From this result is possible to gather information about the value of the coupling constants. In fact, although the value is a combination of the terms in $\Theta_t$ coming from different orders of Lovelock gravity up to order $t$, we have stringent constraint on the value of $\Theta_t$ itself, which should be sufficiently small in order to respect the experimental constraint $c_T^2 \cong 1$ \cite{TheLIGOScientific:2017qsa}.

\section{Scalar perturbations}\label{sec:sc_pert}

In this section we present the first order scalar perturbation equations in a curved FLRW space, in the case $t=3$. We consider the following line element
\begin{equation}
    ds^2 = - \left[1+2\Psi(t,x^k)\right]dt^2 + a(t)^2 \left[1-2\Phi(t,x^k)\right] g_{ij}^{(d-1)}(x^k) \,dx^i dx^j\,,
\end{equation}
where $\Psi$ and $\Phi$ are the Newtonian potentials, whose absolute values is always smaller than unity. The following results are shown in limit $d \rightarrow 4$. 

From the perturbed Einstein equations we can extract the $t-t$ equation
\begin{align}
     3 H \dot{\Phi} - \frac{3 \kappa}{a^2} \Phi &+ \left(Q^2 + 2 J^2 - \frac{3 \kappa}{a^2} \right) \Psi -\frac{\partial_l \partial^l \Phi}{a^2} = -\frac{1}{2 \Gamma M_P^2} \left( \dot{\varphi} \delta \dot{\varphi} + \frac{d V}{d \varphi}\delta \varphi\right)\,,\label{eq:scalar_tt}
\end{align}
the $t-i$ equations
\begin{equation}
    \left(\dot{\Phi} + H \Psi\right) = \frac{\dot{\varphi}}{2 \Gamma M_P^2} \delta \varphi\,,\label{ti}
\end{equation}
the traceless part of the $i-j$ equation
\begin{equation}
    \partial_l \partial^l \Psi - \left(1 + \Theta_3\right) \partial_l \partial^l \Phi = 0\,,\label{traceless_ij}
\end{equation}
and finally the trace of the $i-j$ equation
\begin{align}
    \ddot{\Phi} &+ 3 H \left(1 + \frac{\Theta_3}{3}\right) \dot{\Phi}  + H \dot{\Psi} -\frac{\kappa}{a^2} \left(1+\Theta_3\right) \Phi +\left[J^2 + Q^2 + H^2 \left(1+\Theta_3\right) \right]\Psi - \nonumber
    \\&- \left(1+\Theta_3\right) \frac{\partial_l \partial^l \Phi}{3 a^2} + \frac{\partial_l \partial^l \Psi}{3 a^2} = \frac{1}{2 \Gamma M_P^2} \left( \dot{\varphi} \delta \dot{\varphi} - \frac{d V}{d \varphi}\delta \varphi\right)\,.\label{trace_ij}
\end{align}
In the above equations we neglect an overall factor of $\Gamma$ in all the equations.  Moreover, the definition of $\Theta_3$ is given by Eq. \eqref{eq:Theta} evaluated for $t=3$
\begin{equation}
    \Theta_3 \equiv \frac{8}{\Gamma M_P^2} \left(\alpha + 6 \beta J^2\right) \left(Q^2-J^2\right)\,.
\end{equation}
Note that we can infer a generalization of these equations to any order $t$ of Lovelock gravity, substituting $\Gamma$ with $\Gamma_t$ and $\Omega_3$ with $\Omega_t$.
For completeness we also write the Klein-Gordon equation for the scalar field,
\begin{equation}
    \delta \ddot{\varphi} + 3 H \delta \dot{\varphi} + \frac{d^2 V}{d \varphi^2} \delta \varphi - \frac{\partial_l \partial^l \varphi}{a^2}= \dot{\varphi} \left(3 \dot{\Phi} + \dot{\Psi}\right) - 2 \frac{d V}{d \varphi} \Psi\,,
\end{equation}
which is not affected by the Lovelock term.

We can also obtain an equation for the second derivative of the potential $\Phi$ only. From now we will consider the case $t=2$ for simplicity. We can start from the $t-t$ equation \eqref{eq:scalar_tt}. From the derivative of the $t-i$ equation \eqref{ti}, we can find the term $\dot\varphi\delta \dot{\varphi}$ which we can substituted in the $t-t$ equation. Moreover from the traceless part of the $i-j$ equation we can find the relation between $\Phi$ and $\Psi$ which reads
\begin{equation}
\Psi=\Phi \left[1+\frac{8 \alpha (Q^2-J^2)}{\Gamma M_P^2}\right]\,.  \label{psi}
\end{equation}
Substituting these terms in the $t-t$ equation, and using the background equations to simplify the equation, we obtain
\begin{eqnarray}
\ddot\Phi+&& H \left[
1+\frac{16\alpha (Q^2-J^2)}{M_P^2\Gamma}\right]\dot\Phi\nonumber\\
&&+\left\lbrace 2 (Q^2-J^2)-\frac{2\kappa}{a^2}+\frac{8\alpha}{\Gamma M_P^2}\left[
(Q^2-J^2)^2 + \left(Q^2-\frac{\kappa}{a^2}\right) \dot{H} + H\ddot H
\right]
\right\rbrace\Phi - \frac{\partial_l\partial^{l}\Phi}{a^2} =\nonumber\\
&&= -\frac{\delta\varphi}{M^2_P\Gamma}\left(3H\dot\varphi+\frac{d V}{d \varphi}\right)\,.\label{casalinosebastiani}
\end{eqnarray}
We note that the velocity, i.e. the coefficient of the gradient of $\Phi$, is equal to the speed of light. Therefore, the addition of the regularized Gauss-Bonnet term does not modify the velocity of scalar perturbations with respect to General Relativity.

\section{Inflation}\label{sec:inflation}

In this section we apply the Regularized Lovelock theory to slow-roll inflation, in the flat case $k=0$ and at second order in Lovelock series, i.e. $\beta=0$. We split the discussion in two parts, respectively the scalar and tensor sectors. Our derivation resembles the one in \cite{muk}. Perturbations during slow-roll inflation in the framework of modified theories of gravity are treated in Refs. \cite{Horn1, Horn2, Horn3}
for Hordenski gravity and in Refs. \cite{FR1, FR01, FR02, FR2} for $F(R)$-gravity.

\subsection{Scalar waves during inflation}

By introducing
\begin{equation}
u=\sqrt{\frac{\Gamma}{-\dot H}}\Phi
\,,    \label{uPhi}
\end{equation}
we can rewrite Eq. \eqref{casalinosebastiani} in the flat case as
\begin{equation}
  \ddot u + H\dot u
  +\left[
  2\dot H - \frac{3\ddot H^2}{4 \dot H^2}-\frac{H \ddot H}{2 \dot H}+\frac{\dddot H}{2 \dot H}
  -\frac{4\alpha \dot{H}}{M^2_P \Gamma}
  \left(
  H^2-\frac{3 \dot H}{\Gamma}
  \right)
  \right]u -\frac{\partial_l\partial^l u}{a^2} =0\,,
\end{equation}
where we have used (\ref{ti}) with (\ref{psi}).

We can rewrite the above equation in terms of the conformal time $d\eta=dt/a$ as
\begin{equation}
 \frac{u''}{a^2}-\frac{\partial_l\partial^l u}{a^2}
 +\left[\frac{H^2}{4} +\frac{3 H'}{ 2 a} -\frac{H H''}{2 a H'}-\frac{3 H''^2}{4 a^2 H'^2} + \frac{ H'''}{2 a^2 H'}
 -\frac{4\alpha H'}{M_p^2 \Gamma a} 
 \left(
 H^2 - \frac{3 H'}{a\Gamma}
 \right)
 \right]u = 0\,,
\end{equation}
which can be further simplified in the form
\begin{equation}
u''-\partial_l\partial^l u-\frac{\theta''}{\theta}u= 0\,,   
\label{EOMpert}
\end{equation}
where
\begin{equation}
\theta=\frac{1}{a}\sqrt{\frac{a H^2}{-H'}}
\frac{1}{\sqrt{\Gamma}}
\,.    
\end{equation}
In the short-wavelenght limit $|\theta''/\theta| \ll k^2$, $k$ being the Fourier vector norm associated to the spatial vector of coordinates $x^k$, we obtain the solution
\begin{equation}
u(\eta)=C_0\text{e}^{\pm k\eta}\,,\label{short}    
\end{equation}
where $C_0$ is an integration constant.

On the other hand, 
in the long-wavelength limit $ |\theta''/\theta|\gg k^2$, we have
\begin{equation}
u(\eta)\simeq C_1\theta+ C_2\theta\int\frac{d\eta}{\theta^2}\,, 
\label{long}
\end{equation}
where the $C_1$ mode can be absorbed in the integral and 
\begin{equation}
\int\frac{d\eta}{\theta^2}=
\left(\frac{a}{H}
-\int a^2 d\eta\right)
-\frac{4\alpha H^2}{M_P^2}\left[\frac{a}{H} -\frac{1}{H^2}\int H^2 a^2 d\eta\right]\,.
\end{equation}
When the slow roll approximation is valid the above result can be approximated as
\begin{equation}
\int\frac{d\eta}{\theta^2}\simeq
-\Gamma \frac{ H' }{H^3}\,,
\end{equation}
such that one finally obtains for the long-wavelength limit,
\begin{equation}
u(\eta)=C\,\frac{ \sqrt{-H'/a}}{ H^2} \,\Gamma\,, \label{longslow}   
\end{equation}
where $C$ is an integration constant. Thus, the factor $\Gamma$ contains the correction given by the Regularized Lovelock Gravity to the result of General Relativity.

We also introduce a canonical quantization variable, called $v$, which should be derived from a well-posed Lagrangian derivation. This permits to normalize the vacuum quantum fluctuations generated at the beginning of inflation.
We can infer the action for cosmological perturbations from the equations of motion. In particular, Eq. (\ref{EOMpert}) is a direct consequence of the two relations
\begin{equation}
\partial_l\partial^l u= z\left(\frac{v}{z}\right)'\,,\qquad v=\theta \left(\frac{u}{\theta}\right)'\,,\label{relations}
\end{equation}
where
$
z \equiv 1/\theta
$.
For comparison with the set of field equations presented in the Sec. \ref{sec:sc_pert}, we should consider $v=a[\delta\varphi+(a\varphi'/a')\Phi]/\sqrt{2M_P^2}$.

The quadratic action for perturbations in terms of $v$ is
\begin{equation}
    S_v=\frac{1}{2}\int d\eta \,d^3x \left(v'^2+v\partial_l\partial^l v+\frac{z''}{z} v^2\right) \,,   
\end{equation}
with the associated equation of motion,
\begin{equation}
    v''-\partial_l\partial^l v -\frac{z''}{z}v=0\,.
\end{equation}
After decomposing $v$ in Fourier modes $v_k$ and introducing the Bunch-Davies vacuum state as the boundary condition of the solution, such that 
$v_k(\eta_i)=1/\sqrt{k}$ and $v_k'(\eta_i)=i\sqrt{k}$ at the reference (conformal) time $\eta_i$, we can use the Eq.s (\ref{short}) and (\ref{relations}) to derive the corresponding modes $u_k$ in
the short-wavelength limit. We obtain
\begin{equation}
u_k (\eta)=\frac{-i}{k^{3/2}}\text{e}^{i k (\eta-\eta_i)}\,. \label{u1}
\end{equation}
After a Fourier mode enters the long-wavelength regime, its evolution is described by Eq. (\ref{long}), namely
\begin{align}
u_k(\eta) &\simeq A_k\theta\int\frac{d\eta}{\theta^2}=\nonumber\\
&=\frac{A_k}{\sqrt{\Gamma}}\sqrt{\frac{a }{-H'}}\left[\left(1
-\frac{H}{a}\int a^2 d\eta\right)
-\frac{4\alpha H^2}{M_P^2}\left(1 -\frac{1}{a H}\int H^2 a^2 d\eta\right)\right]\,,
\label{u2}
\end{align}
where the amplitude $A_k$ can be fixed by comparing (\ref{u1}) and (\ref{u2}) at the time when the Fourier mode crosses the Hubble horizon. Thus, using the slow-roll approximation derived in (\ref{longslow}) for the modes $u_k$, after the identification of $C=A_k$ we find
\begin{equation}
A_k\simeq \left(-\frac{i}{k^{3/2}}\frac{H^2}{\sqrt{-H'/a}}\frac{1}{\Gamma}\right)_{k\simeq H a}\,.    
\end{equation}
Thus, taking into account (\ref{uPhi}), the scale-invariant power spectrum of long-wavelength scalar perturbations is
\begin{eqnarray}
\delta_\Phi^2&&=\frac{1}{8\pi^2 M_P^2}\left(\frac{-H'}{a\Gamma}\right) |u_k|^2 k^3=\nonumber\\
&&\hspace{-1cm}=
\frac{1}{8\pi^2 M_P^2}\left|\frac{1}{\Gamma^2}
\left(\frac{ a H^4}{-H'}\right)\right|_{k\simeq Ha}\frac{1}{\Gamma^2}
\left[\left(1
-\frac{H}{a}\int a^2 d\eta\right)
-\frac{4\alpha H^2}{M_P^2}\left(1 -\frac{1}{a H}\int H^2 a^2 d\eta\right)\right]^2\,,
\end{eqnarray}
which is valid for perturbations with wavelength $\lambda\equiv2\pi/k>1/(H a)$. 
Note that, contrarily to the case of General Relativity, the scalar perturbations are not completely frozen during the radiation dominated era before they re-enter the horizon. In fact, a time dependence is preserved, as we can see from the following result. Using a scale factor time dependence $a(t)\propto t^{1/2}$, i.e. the scale factor time evolution in a radiation dominated period, we obtain the following
\begin{equation}
 \delta_\Phi^2=   
 \frac{1}{8\pi^2 M_P^2}\left|\frac{1}{\Gamma^2}
\left(\frac{ a H^4}{-H'}\right)\right|_{k\simeq Ha}\frac{4}{9}
\left(1-\frac{12 \alpha H^2}{M_P^2}\right)^2\left(1+\frac{4\alpha H^2}{M_P^2}\right)^{-2}\,,
\label{powerspectrumscalar}
\end{equation}
and only if $\alpha=0$ we recover the constant result of General Relativity.

\subsection{Gravitational waves during inflation}

In this section we study the tensor perturbations described by Eq. (\ref{tp}) in the context of inflation. Also in this case, we consider $\beta=0$ and $k=0$. After the decomposition of 
$h_{ij}$ in Fourier modes $h_k$, and introducing a the new variable 
\begin{equation}
y_k \equiv \Gamma\,a \,h_k\,,
\end{equation}
we obtain, in terms of the conformal time,
\begin{align}
y''_k&-\left(1+\frac{8\alpha H'}{ \Gamma a M_P^2}\right)\partial_l\partial^l y_k -\nonumber\\
&-\left[
2a^2H^2+H'+\frac{\alpha}{M_P^2\Gamma}
\left(
8 H'H^2 a
-4H H''
-\frac{4H'^2 }{\Gamma}
\right)
\right]
y_k
=0 
\,.
\end{align}
Note that this equation of motion can be obtained from a well posed action derivation. 

On the de Sitter space-time with $H$ constant and $a=-1/(H\eta)$ we simply have
\begin{equation}
y_k''-\partial_l\partial^l y_k-\frac{2}{\eta^2}\,y_k=0\,. \end{equation}
The above equation form coincides with the one of General Relativity. In fact the Lovelock corrections vanish as they depend on the derivatives of $H$, which are null on de Sitter.

Considering the standard initial conditions given by $y_k(\eta_i)=1/\sqrt{k}$ and $y_k(\eta_i)=i\sqrt{k}$ with $ k|\eta_i|\gg 1$, we obtain the solution
\begin{equation}
y_k= \frac{1}{\sqrt{k}}\left(1+\frac{i}{k\eta}\right)\text{e}^{ik(\eta-\eta_i)}\,,
\end{equation}
such that the power spectrum of tensor perturbations is given by
\begin{equation}
\delta_h^2=\frac{1}{\pi^2 M_P^2}\frac{|y_k|^2 k^3}{a^2\Gamma^2}= \frac{1}{\pi^2 M_P^2}\frac{H^2}{\Gamma^2}(1+k^2\eta^2)\,,
\quad \frac{k}{a}\gg\frac{H\eta}{\eta_i}\,.
\end{equation} 
On the other hand, long-wavelength tensor mode perturbations with 
$\lambda \, a= 2\pi a/k\gg (\eta_i/\eta)/H$ have a flat spectrum with an amplitude proportional to $H/\Gamma$. Since during inflation $H$ changes slowly, the power spectrum of long-wavelength tensor perturbations is fixed at the time when these perturbations cross the horizon, namely
\begin{equation}
\delta_h^2= \frac{1}{\pi^2 M_P^2}\frac{H^2}{\Gamma^2}|_{k\simeq Ha}\,.    \label{deltah}
\end{equation}
We note that the tensor perturbations power spectrum gains some corrections from the Regularized Lovelock Gravity through the factor $\Gamma$.

\subsection{Spectral indices and tensor-to-scalar power spectra ratio}

In this subsection we show the equations for the spectral indices of scalar and tensor perturbations, and for the tensor-to-scalar power spectra ratio. The results are given in terms of the cosmological time. 

The expression for the spectral index of the scalar perturbations is derived from Eq. (\ref{powerspectrumscalar}). Using the definition, we obtain
\begin{equation}
n_s-1\equiv \frac{d \delta_\Phi^2}{d\log k}
=4\frac{\dot H}{H^2}-\frac{\ddot H}{H\dot H}-\frac{16\alpha \dot H}{M^2_P\Gamma}|_{k\simeq Ha}\,,
\end{equation}
which can be checked against the experimental results \cite{Planck} once the scalar field potential is fixed. The spectral index for the tensor perturbations is defined as
\begin{equation}
n_T \equiv  \frac{d \delta_T^2}{d\log k}=\frac{2 \dot H}{H^2}  -\frac{16\alpha \dot H}{M^2_P\Gamma}|_{k\simeq Ha}\,,
\end{equation}
where we used Eq. (\ref{deltah}). 

Finally, the tensor-to-scalar spectral ratio is
\begin{equation}
r\equiv 2\frac{\delta_{\Phi}^2}{\delta_h^2}=16\left(-\frac{\dot H}{H^2}\right)|_{k\simeq Ha}\,,
\end{equation}
where the two polarizations of the tensor modes are taken into account.

Therefore, the spectral indices results reduce to the ones of General Relativity in the limit $\alpha\rightarrow 0$, while the form of the tensor-to-scalar spectra ratio is not affected by Regularized Lovelock Gravity corrections. 

\section{Conclusions}\label{sec:conclusions}

The Regularized Lovelock Gravity theory permits to circumvent the Lovelock theorem. In this theory, the action is formulated in $d$ dimensions and by adding a particular combination of the curvature tensors, given by Lovelock theorem, to the Hilbert-Einstein action of General Relativity. Then, by using the dimensional regularization and by taking the limit $d\rightarrow 4$ of the field equations, these additional tensors provide non trivial contributions which would be vanishing in four dimension without  regularization. The first Lovelock correction is given by the Gauss-Bonnet term, and was firstly introduced by Tomozawa \cite{Tomozawa:2011gp} to deal with finite one-loop quantum corrections to Einstein gravity, and recently it has been re-proposed in \cite{Glavan:2019inb}.

In this work, we have studied the first order perturbations in Regularized four dimensional Lovelock Gravity on a general non-flat FLRW background, in the presence of a canonical scalar field. In particular, we have derived the scalar and tensor perturbations up to the cubic order of Lovelock gravity, extending the results of Glavan and Lin \cite{Glavan:2019inb} which focused on the flat FLRW background and on the quadratic Regularized Lovelock Gravity. We found that this theory modifies the velocity of tensor waves with respect to the well known result $c_T^2=1$ of General Relativity. These corrections may  be taken arbitrarily small, since they depend explicitly on the Lovelock parameters $\alpha$ and $\beta$. Moreover, we were able to compute an expression for the velocity of gravitational waves at a generic order, showing that higher order Lovelock terms introduce additional corrections. 
We also note that the speed of gravitational waves is highly constrained by the recent LIGO/VIRGO \cite{Ligo1} experiment (and see \cite{Ligo2,Ligo3,Ligo4,Ligo5,Ligo6} for the impact on modified gravity theories) and this result yields the possibility of future detection of the correction terms in experiments with higher sensitivity such as the LISA experiment. 
On the other hand, we have found that the velocity of scalar perturbations corresponds to the velocity of light at any order in Lovelock gravity.

In the second part of our work, we have studied the theory of cosmological perturbations in a slow-roll scalar field inflation, on the flat FLRW background. We have found that the spectral indices acquire some corrections with respect to General Relativity results. On the contrary, the tensor-to-scalar spectra ratio assumes the same form.

\subsection*{Acknowledgments}

This work has been partially performed using the software \texttt{xAct} \cite{xAct} and \texttt{xPand} \cite{Pitrou:2013hga}. A.\ C.\ acknowledges the financial support of the Italian Ministry of Instruction, University and Research (MIUR) for his Doctoral studies. The authors thank Aimeric Coll\'eaux, Massimiliano Rinaldi, Silvia Vicentini and Sergio Zerbini for the useful discussions.

{\pagestyle{empty}\cleardoublepage}
  
\end{document}